\documentclass[conference]{IEEEtran}
 
\usepackage{cite}
\ifCLASSINFOpdf
   \usepackage[pdftex]{graphicx}
   \graphicspath{{../pdf/}{../jpeg/}}
   \DeclareGraphicsExtensions{.pdf,.jpeg,.png}
\else
\fi
\usepackage{color}
\usepackage{amsmath}
\usepackage{algorithmic}
\usepackage{array}
\usepackage{url}

\usepackage[]{fancyhdr} % 
\newcommand{\changefont}{\fontsize{9}{9}\selectfont}
\ifCLASSOPTIONcompsoc
 \usepackage[caption=false,font=normalsize,labelfont=sf,textfont=sf]{subfig}
\else
 \usepackage[caption=false,font=footnotesize]{subfig}
\fi

\usepackage{fixltx2e}

\usepackage{stfloats}
\IEEEoverridecommandlockouts
\usepackage{xcolor}

\begin{document}

% Titles are generally capitalized except for words such as a, 、【】an, and, as,
% \title{Selection of Cyberattack Frequency Based on Eigenanalysis and Transfer Function}
% \title{Demand-side Energy Storage Devices as a Potential Cyberattack Resource}

\title{Cyber-Physical Attack Leveraging Subsynchronous Resonance}

\author{\IEEEauthorblockN{Bosong Li, Baosen Zhang, Daniel S. Kirschen}
\IEEEauthorblockA{Department of Electrical and Computer Engineering\\University of Washington\\
}
}


%Telephone: (800) 555--1212, Fax: (888) 555--1212}

% <-this % stops a space


%\lhead{11TH BULK POWER SYSTEMS DYNAMICS AND CONTROL SYMPOSIUM, JULY 25-30, 2022, BANFF, CANADA}
%\rhead{1}

%\fontfamily{phv}\fontseries{b}\fontsize{9}{11}\selectfont

\maketitle
\thispagestyle{fancy}
\pagestyle{fancy}

%\thispagestyle{fancy}
%\pagestyle{fancy}

\begin{abstract}
This paper discusses how a cyber attack could take advantage of torsional resonances in the shaft of turbo-generators to inflict severe physical damage to a power system. If attackers were able to take over the control of a battery energy storage device, they could modulate the injection of this device at a frequency that matches one of the sub-synchronous resonance frequencies of a generator. Small changes in injection might be sufficient to excite one of these mechanical resonances, resulting in metal fatigue and ultimately a catastrophic failure in the shaft of the generator. Using a state-space model of the electro-mechanical system, the paper develops transfer functions linking the magnitude of the malicious injections to the magnitude of oscillations in the speed and angle of the various masses connected to the shaft. Numerical results from a two-area power system demonstrate the existence of vulnerable frequencies and show that damaging mechanical oscillations can be triggered without causing easily detectable signals at the generator terminals.

\end{abstract}



% \fi
\IEEEpeerreviewmaketitle

\section{Introduction}
While a cyber attack \cite{8672483, 8680713} \cite{wang2019review, 8260948}\cite{zheng2020security,mousavian2017risk} cripples a power system until the malicious software has been expurgated, a physical attack that destroys a major piece of equipment can take months to repair. To carry out a physical attack, the adversary must get in close proximity to the target equipment, which is expensive and difficult to carry out without detection. On the other hand, a cyber-physical attack does not require the malicious actor to get in close proximity of the target equipment. Instead, the attacker infiltrates some aspect of the  control infrastructure of the power system and manipulates it to create physical damage \cite{nguyen2020electric}.  Idaho National Laboratory demonstrated the feasibility of this type of attack by taking over the protection system of a generator and manipulating its synchronization until the generator self-destructed \cite{zeller2011myth}.

To avoid detection and countermeasures, malicious manipulations of the control system should remain small. It is therefore important to explore how an attacker could use resonances to amplify their effects. Because their output can be modulated at high frequency, battery energy storage systems (BESS) represent an ideal vector for this type of attack. Furthermore, because the number of BESS deployed in power systems is increasing rapidly, the attack surface is growing \cite{assante2010high} \cite{tejada2019review} \cite{glenn2016cyber}.

This paper explores how a malicious actor could damage large turbo-generators by taking over the control of a BESS and using it to inject small amounts of power at frequencies corresponding to the torsional sub-synchronous resonance (SSR) frequencies in the  shaft of some generators. SSR is a condition of the electric power system where with a turbine generator exchanges energy with the rest of the system at one or more of the natural frequencies of the combined system below the synchronous frequency of the system \cite{ieee1981proposed}. While sub-synchronous oscillations between the various masses connected to the shaft of the generator typically remain small, they cause  metal fatigue and can over time lead to a catastrophic failure of this shaft \cite{anderson1999subsynchronous}.

Physical and control countermeasures can be taken to avoid SSR \cite{kundur2007power}. Capacitor compensation can be added to the system, but the investment cost of such infrastructure measures is high. Protective relays can also be used to detect oscillations at the generator terminals. However, as this paper will show, it is possible to induce mechanical oscillations that are hard to detect at the generator terminals.

To demonstrate the feasibility of cyber-physical attacks that leverage sub-synchronous resonances, this paper develops a state-space model of the combined electromechanical system. From this state space model, we derive transfer functions linking the magnitude of the disturbances created by the subverted BESS and the angular frequency of the various masses connected to the shaft of a generator. These transfer functions exhibit resonance frequencies that an attacker could target. A relatively low-power BESS could therefore trigger sub-synchronous resonances that could ultimately destroy large generators while remaining hard to detect.

The remainder of the paper is organized as follows:
Section \ref{S2} describes a state-space model of the system dynamics that combines the swing equations of the generator, the mechanical characteristics of the shaft system, as well as the power flow equations.
Section \ref{S3} derives transfer functions relating the BESS injections to the angular frequency and position of the various masses connected to the generator shaft. 
Section \ref{S4} describes numerical studies that illustrate the frequency-domain analysis and correlate it with time-domain simulations. Section \ref{SecAdd} discusses possible countermeasures. Section \ref{S5} concludes and discusses further work.

\section{Model of System Dynamics}
\label{S2}
\subsection{Notation}
In this paper, the rated angular velocity in electrical rad/s is denoted by $\omega_0 = 2\pi f_0$, where frequency $f_0 = 60$ Hz.
Assuming the number of field poles $p_f = 2$, we employ $\omega_{0m}$ as the rated angular velocity in mechanical rad/s, and $\omega_{0m} = (2/p_f)\omega_0 = 377$ rad/s.

% \todo{Do we need to specify the number of poles?}
% \sectodo{Since we mention $\omega_{0m}=\frac{2}{pf}\omega_0$,I specified the value of $p_f$, of we can just say $\omega_{0m}=377rad/s$ and left the $p_f$ part out.}

\begin{figure}[h]
\centering
\includegraphics[width=0.9\linewidth]{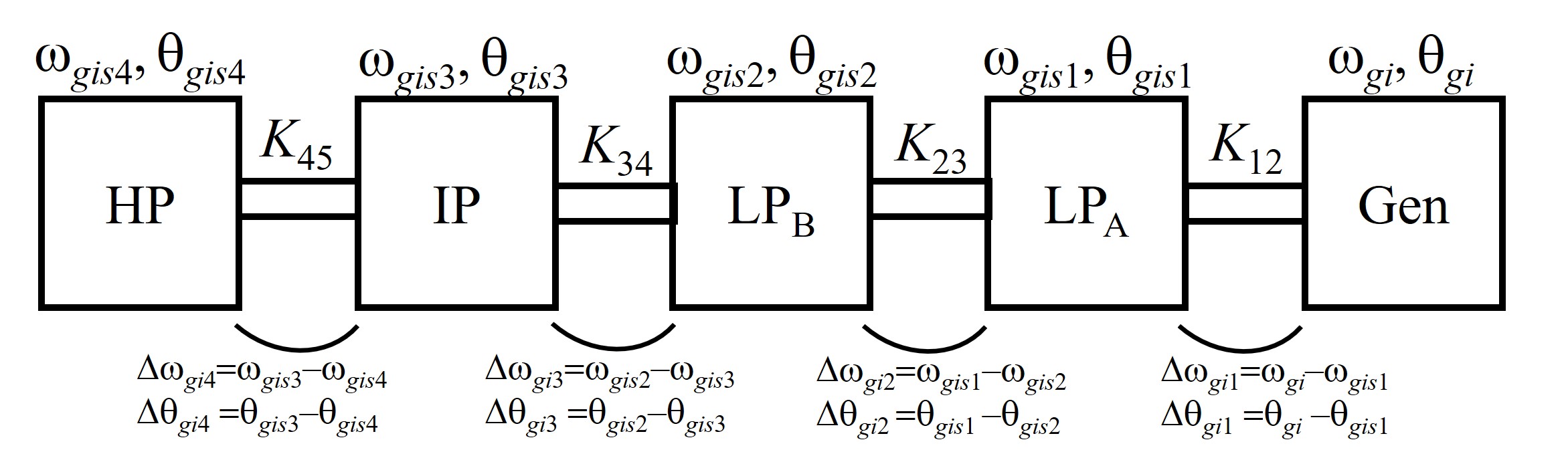} 
\caption{Five-Mass Torsional System.}
\label{fig_sh}
\end{figure}

Each generator contains a five-mass torsional system.
This paper denotes the speed and angle deviation of each rotor from the steady-state values respectively with $\omega_g, \omega_{s1}, \omega_{s2}, \omega_{s3}, \omega_{s4}$ and $\theta_g, \theta_{s1}, \theta_{s2}, \theta_{s3}, \theta_{s4}$.
The subscription $gi$ represents the terminal of generator $i$, while the subscriptions $s1-s4$ represent the other turbine sections.
We further define $\Delta \omega$ and $\Delta\theta$ as the speed and angle difference between two adjacent masses connecting to the same shaft.
A torsional system of generator $i$ is presented in Fig. \ref{fig_sh} with the variables illustrated, 
where the shaft between two adjacent masses is denoted using a double subscript as $12,23,34$ and $45$.

In the following formulation, the variables and parameters are employed with their per-unit values if not specially mentioned.

\subsection{State-Space Model}
We investigate the dynamics of generators considering the swing equations of the generator rotor together with the torsional characteristics of the shaft system shown below in
\eqref{swing}-\eqref{sw2} and \eqref{shaft01}-\eqref{shaft42}.
\begin{align}
\label{swing}
    & \dot{\omega}_g = \frac{\omega_{0m}}{2H_g}~(P_M-P_e-D_g \cdot \omega_g)\\ 
\label{sw2}
    &\dot{\theta}_g = \omega_g
\end{align}

\begin{align}
\label{shaft01}
&\dot{\omega}_{g} = \frac{\omega_{0m}}{2H_g}~\big(K_{12}(\theta_{s2}-\theta_g)-D_g\cdot \omega_g-P_e\big)\\ 
\label{shaft11}
&\dot{\omega}_{s1} = \frac{\omega_{0m}}{2H_{s1}}~\big(K_{23}(\theta_{s3} - \theta_{s2}) - K_{12}(\theta_{s2} - \theta_g) \\ \nonumber
&~~~~~~~~- D_{s1}\cdot \omega_{s1} + P_{s1}\big) \\
\label{shaft21}
&\dot{\omega}_{s2} = \frac{\omega_{0m}}{2H_{s2}}~\big(K_{23}(\theta_{s3} - \theta_{s2}) - K_{12}(\theta_{s2} - \theta_{s1}) \\ \nonumber
&~~~~~~~~- D_{s2}\cdot \omega_{s2} + P_{s2}\big) \\
\label{shaft31}
&\dot{\omega}_{s3} = \frac{\omega_{0m}}{2H_{s3}}~\big(K_{34}(\theta_{s4} - \theta_{s3}) - K_{23}(\theta_{s3} - \theta_{s2}) \\ \nonumber
&~~~~~~~~- D_{s3}\cdot \omega_{s3} + P_{s3}\big) \\
\label{shaft41}
&\dot{\omega}_{s4} = \frac{\omega_{0m}}{2H_{s4}}~\big(-K_{34}(\theta_{s4}-\theta_{s3})-D_{s4}\cdot \omega_{s4}-P_{s4}\big)\\ 
\label{shaft12}
&\dot{\theta}_{s1} = \omega_{s1}\\
\label{shaft22}
&\dot{\theta}_{s2} = \omega_{s2}\\
\label{shaft32}
&\dot{\theta}_{s3} = \omega_{s3}\\
\label{shaft42}
&\dot{\theta}_{s4} = \omega_{s4}
\end{align}
% \todo{explain some of the parameters here, like what $H$ is.}

\textcolor{black}{In \eqref{shaft01}-\eqref{shaft41}, parameters $D$ and $H$ are respectively the damping coefficient and inertia constant of each rotor section. Parameter $K$ is the shaft stiffness.}
Note that \eqref{swing} and \eqref{shaft01} illustrate the dynamics of the generator rotor in different forms. 
Furthermore, the mechanical torque on the generator rotor equals the electrical torque in the steady state, i.e. $P_M = P_e$.
We can therefore combine \eqref{swing} and \eqref{shaft01} as \eqref{shaftsub}.
\begin{equation}
\label{shaftsub}
  \dot{\omega}_{g} = \frac{\omega_{0m}}{2H_g}~\big(K_{12}(\theta_{s2}-\theta_g)-D_g\cdot \omega_g-P_M\big) 
\end{equation}
% \todo{We are not doing control, so just dynamics, not control dynamics.}
Hence the dynamics of a generator considering the shaft system is fully described by \eqref{sw2}, \eqref{shaft11}-\eqref{shaft42}, and \eqref{shaftsub}.
In a power system containing $n$ generators, we take the rotor speed $\boldsymbol{\omega}$ and angular displacement $\boldsymbol{\theta}$ of each generator as state variable $\mathbf{x}=[\boldsymbol{\omega}^T,\boldsymbol{\theta}^T]^T$, 
where $\boldsymbol{\omega}$ and $\boldsymbol{\theta}$ are both vectors of length $5n$ illustrated in details as:
\begin{align}
&\boldsymbol{\omega} = [\underbrace{\omega_{g1},\omega_{g2},...,\omega_{gn}}_n,\underbrace{\omega_{g1s1},...,\omega_{g1s4},...,\omega_{gns1},...,\omega_{gns4}}_{4n}
    ]^T  \\
&\boldsymbol{\theta} = [\underbrace{\theta_{g1},\theta_{g2},...,\theta_{gn}}_n,\underbrace{\theta_{g1s1},...,\theta_{g1s4},...,\theta_{gns1},...,\theta_{gns4}}_{4n}
    ]^T    
\end{align}

In the steady state, the mechanical power $P_M$ of the generator rotor in \eqref{shaftsub} equals the electrical power.
By substituting the load bus angular displacement in the power flow and power balance equations with state variables, we write the mechanical power using the angular displacement vector $\boldsymbol{\theta}$ and load vector $\mathbf{L}$: 
% \todo{is angular displacement vector or just angle}    
\begin{equation}
\label{pm}
    \mathbf{P_M} = \mathbf{P_e} = \mathbf{A_e}\boldsymbol{\theta}+\mathbf{B_e}\mathbf{L}
\end{equation}
where $\mathbf{P_M} = [P_{Mg1},P_{Mg2},...,P_{Mgn}]^T$. 
The matrices $\mathbf{A_e}$ and $\mathbf{B_e}$ depend on the system admittance matrix and topology.
\textcolor{black}{Since the focus of this paper is on the oscillations between different masses in the generator, we take a load bus to be the slack bus to emphasis generators internal dynamics.}
% \todo{This sentence is not clear. Try something like: Since the focus of this paper is on the oscillations between different masses in the generator, we take a load bus to be the slack bus to emphasis generators internal dynamics.}

This paper further denotes the input power on each mass of the generator shaft system with a 4-dimensional vector $\mathbf{P_{gi}} = [P_{gis1},P_{gis2},P_{gis3},P_{gis4}]^T = P_{Mgi} \cdot \mathbf{B_f}_{gi} $,
% where vector $\mathbf{B_f}_{gi}$ is the power fraction coefficient of each mass. 
\textcolor{black}{where the coefficient vector $\mathbf{B_f}_{gi}$ contains the fraction of the total turbine power generated by each turbine in the steady-state.}
% \todo{is power fraction coefficients a standard term?} 
% \sectodo{The [Kundur] book referred to this parameter in a table as 'power fraction', I change the term to its explanation to be clearer.}
Following the pattern of state variables $\boldsymbol{\omega}$ and $\boldsymbol{\theta}$, the input power vector $\mathbf{P_I}=[\mathbf{P_M}^T,\mathbf{P_{g1}}^T,...,\mathbf{P_{gn}^T}]^T$ is defined as:

\begin{equation}
\label{inp}
    \mathbf{P_I} 
    = \mathbf{B_I}\cdot \mathbf{P_M} = \mathbf{B_I}\mathbf{A_e}\boldsymbol{\theta}+\mathbf{B_I}\mathbf{B_e}\mathbf{L}
\end{equation}
% The matrix $\mathbf{B_I}$ is a $5n \times n$ matrix defined as:
\begin{equation}
    \mathbf{B_I} = 
    \left[\begin{smallmatrix}
    \mathbf{I_{n\times n}} \\ \mathbf{B_F}
    \end{smallmatrix}\right]
\end{equation}
\begin{equation}
    \mathbf{B_F} = 
    \left[\begin{smallmatrix}
    \mathbf{B_f}_{g1} & & & \\ & \mathbf{B_f}_{g2}& & \\ & & \ddots & \\ & & &\mathbf{B_f}_{gn}
    \end{smallmatrix}\right]
\end{equation}
where $\mathbf{I_{n\times n}}$ is an $n$ by $n$ identity matrix.
The $4n \times n$-dimensional coefficient matrix $\mathbf{B_F}$ is constructed by matrices $\mathbf{B_f}_{g1}, \mathbf{B_f}_{g2},..., \mathbf{B_f}_{gn}$ indicating how much input power each turbine provides in a generator. 
% \todo{again power fraction matrices reads weird}

We therefore build the state-space model of the whole power system as follows:
\begin{equation}
\label{con1}
    \left[ \begin{array}{c}
    \dot{\boldsymbol{\omega}} \\ \dot{\boldsymbol{\theta}}
    \end{array} \right] = 
    \left[ \begin{array}{cc}
    \mathbf{A_{11}} & \mathbf{A_{12}}\\ \mathbf{A_{21}} & \mathbf{A_{22}} 
    \end{array} \right] 
    \left[ \begin{array}{c}
    \boldsymbol{\omega} \\ \boldsymbol{\theta}
    \end{array} \right] 
    + \mathbf{B} \cdot\mathbf{L} 
\end{equation}
The system matrix $\boldsymbol{A}$ consists of four parts, where:
\begin{align}
&\mathbf{A_{11}} \!=\! -\frac{\omega_{0m}}{2}\mathbf{diag}{(-\frac{D_g}{H_g},\! -\frac{D_{s1}}{H_{s1}},\!-\frac{D_{s2}}{H_{s2}},\!-\frac{D_{s3}}{H_{s3}},\!-\frac{D_{s4}}{H_{s4}})}  \\
&\mathbf{A_{12}} \!\!=\!\!\mathbf{B_I}\!\mathbf{A_e} \!\!+\!\! \frac{\omega_{0m}}{2}
\!\!\!\left[
{\small \begin{matrix}
\begin{smallmatrix}
    -\frac{K_{12}}{H_{g}} & \frac{K_{12}}{H_g} & & & \\
    \frac{K_{12}}{H_{s2}} & -\frac{K_{12}+K_{K23}}{H_{s2}} & \frac{K_{23}}{H_{s2}} & & \\
      & \frac{K_{23}}{H_{s3}} & -\frac{K_{23}+K_{34}}{H_{s3}} & \frac{K_{34}}{H_{s3}} & \\
      & & \frac{K_{34}}{H_{s4}} & -\frac{K_{34}+K_{45}}{H_{s4}} &\frac{K_{45}}{H_{s4}} \\
      & & & \frac{K_{45}}{H_{s5}} & -\frac{K_{45}}{H_{s5}} \\
\end{smallmatrix}
\end{matrix}}
\right]
\end{align}
The sub-matrix $\mathbf{A_{21}} \!=\!\mathbf{I_{5n\times5n}}$ is an identity matrix, while sub-matrix $\mathbf{A_{22}} \!=\!\mathbf{0}$ is a zero matrix.
The input matrix $\mathbf{B}=\mathbf{B_I}\mathbf{B_e}$ according to \eqref{inp}.

\section{Transfer Function Magnitude Analysis}
\label{S3}
With access to the BESS on a load bus, potential attackers can inject oscillatory signals into the system, illustrated mathematically as \eqref{attack}, where the attack signal $\Delta L$ is a square-wave signal. 
% \todo{Explain why the attacker would choose oscillatory signals: 1) it's a storage 2) it is in some sense the worst-case attack.}
% \sectodo{The attack signal we use is a square wave rather than a sine wave as a result of our MPC solution with the objective of max(generator oscillation). Below is what I think:}
\textcolor{black}{
This paper assumes a square-wave attack signal because it is created by an energy storage device by constantly changing between the charging and discharging states.
Moreover, we built a model predictive control (MPC) problem to maximize the oscillation of the generator terminal variables, and the optimal solution to this problem was a square wave.
% which is also a simulation of the rapid disconnection and recorrection in the Aurora Test to damage a generator  \cite{zeller2011myth}. 
}

\begin{equation}
\label{attack}
    \dot{\mathbf{x}} = 
    \left[ \begin{array}{cc}
    \mathbf{A_{11}} & \mathbf{A_{12}}\\ \mathbf{A_{21}} & \mathbf{A_{22}} 
    \end{array} \right]\mathbf{x} 
    + \mathbf{B} (\mathbf{L} +\Delta\mathbf{L})
\end{equation}
% where $\Delta\mathbf{L}$ is the injected attack.
All the elements of the attack signal $\Delta\mathbf{L}$ are 0 except for the element corresponding to the load bus where the malicious energy storage device resides. 
We assume the oscillatory signal magnitude is 1 p.u. of the base power
considering the limited power and capacity of the attack energy storage device. 
% \todo{This depends on what the base unit is. Is this assumption needed? Can we just say we normalize such that the attack magnitude is 1?}
% \sectodo{I mention the variables are in p.u. if not specially mentioned in the Notation part, will that make the 1 p.u. here make sense?}
The only factor that affects the system oscillation is, therefore, the frequency of input signal $\Delta\mathbf{L}$.
% \todo{Again, it's important to explain why a sinusoid would be used.}
To explore the impact of the injected oscillatory signal frequency on each rotor of the generators in the system, we investigate two output signals denoted as \eqref{output1} and \eqref{output2}. 
% \todo{The important point here to make is that we want to compare these outputs. And say what we will see.}
\textcolor{black}{To explore the impact of the oscillatory signal injection on the generator terminal and the torsional system, we define the following output: }
\begin{align}
\label{output1}
&\mathbf{y_1} = 
\left[ \begin{array}{c}
\boldsymbol{\omega} \\ \boldsymbol{\theta}
\end{array} \right] = \mathbf{C_1}\mathbf{x}+\mathbf{D_1}(\mathbf{L} +\Delta\mathbf{L})\\
\label{output2}
&\mathbf{y_2} = 
\left[ \begin{array}{c}
\Delta\boldsymbol{\omega} \\ \Delta\boldsymbol{\theta}
\end{array} \right] = \mathbf{C_2}\mathbf{x}+\mathbf{D_2}(\mathbf{L} +\Delta\mathbf{L})
\end{align}
% \todo{We are not really comparing vectors, rather some component in the vector. The phrase "compare $y_1$ to $y_2$ is the not the most helpful. Try just saying the torsional variables and generator terminal variables }
% \sectodo{The following content in this section related to $\mathbf{y_1}$ and $\mathbf{y_2}$ is modified accordingly.}
\textcolor{black}{where output $\mathbf{y_1}$ is the generator terminal variables.
The output $\mathbf{y_2}$ contains torsional variables showing the speed and angle differences between the rotors connecting to the same shaft.
Output $\mathbf{y_2}$ is further explained in \eqref{tord1}-\eqref{tord4}.}
\begin{align}
\label{tord1}
&\Delta \boldsymbol{\omega} = [\Delta \boldsymbol{\omega}_{g1},\Delta \boldsymbol{\omega}_{g2},...,\Delta \boldsymbol{\omega}_{gn}]^T \\
\label{tord2}
&\Delta \boldsymbol{\theta} = [\Delta \boldsymbol{\theta}_{g1},\Delta \boldsymbol{\theta}_{g2},...,\Delta \boldsymbol{\theta}_{gn}]^T
\end{align}
For generator $i$, $\Delta \boldsymbol{\omega}_{gi}$ and $\Delta \boldsymbol{\theta}_{gi}$ are both 4-dimensional vectors:
\begin{align}
\label{tord3}
&\Delta \boldsymbol{\omega_{gi}} \!\!=\!\! [\omega_{gi}\!\!-\!\!\omega_{gis1},\omega_{gis1}\!\!-\!\!\omega_{gis2},\omega_{gis2}\!\!-\!\!\omega_{gis3},\omega_{gis3}\!\!-\!\!\omega_{gis4}]^T\\ 
\label{tord4}
&\Delta \boldsymbol{\theta_{gi}} \!\!=\!\! [\theta_{gi}\!\!-\!\!\theta_{gis1},\theta_{gis1}\!\!-\!\!\theta_{gis2},\theta_{gis2}\!\!-\!\!\theta_{gis3},\theta_{gis3}\!\!-\!\!\theta_{gis4}]^T
\end{align}
We analyze the magnitudes of the transfer functions between the input signal $\Delta \mathbf{L}$ and the output signals in \eqref{output1} and \eqref{output2} over the frequency spectrum from 0Hz to 60Hz.
The magnitudes of the transfer functions $|\Gamma_1|_j$ and $|\Gamma_2|_k$ correspond respectively to the $j th$ generator terminal variable and the $k th$ torsional variable of a generator. 

\begin{align}
\label{y1}
&|y_1(j\omega_a)|_j = |\Gamma_1(j\omega_a)|_j \cdot |\Delta L(j\omega_a)|\\
\label{y2}
&|y_2(j\omega_a)|_k = |\Gamma_2(j\omega_a)|_k \cdot |\Delta L(j\omega_a)|
\end{align}
Equations \eqref{y1} and \eqref{y2} show that with the same oscillatory input signal, the oscillation magnitudes of the generator terminal variables as well as the torsional variables are determined by the magnitude of the corresponding transfer function.
With an oscillatory input signal at frequency $\omega_a$, if the magnitude $|\Gamma_2(j\omega_a)|_k$ is higher than $|\Gamma_1(j\omega_a)|_j$ as shown in \eqref{con}, 
the internal torsional system of the generator suffers a more severe oscillation compared to the generator terminal.
\begin{equation}
\label{con}
    |\Gamma_2(j\omega_a)|_k > |\Gamma_1(j\omega_a)|_j, j,k \in the~same~generator
\end{equation}
Additionally, the protection system is usually designed to prevent the generator terminal rotor from significant oscillation, i.e., damping the frequencies that lead to high $|\Gamma_1(j\omega_a)|_i$ values.
The input oscillation frequencies denoted in \eqref{con} therefore remain commonly neglected.
A ratio parameter $R_M$ is further defined in \eqref{rdef} as the ratio between the transfer function magnitudes of the torsional output and the terminal output.
\begin{equation}
\label{rdef}
    R_M(\omega_a) = \frac{|\Gamma_2(j\omega_a)|_k}{|\Gamma_1(j\omega_a)|_i}, i,k \in the~same~generator
\end{equation}
The higher the ratio $R_M$ is, the more severe the oscillation inside the generator shaft system occurs while the measurements at the generator terminal remain close to the steady-state values.
If the attack frequency $\omega_a$ in \eqref{rdef} is lower than the nominal frequency, i.e. $\omega_a<\omega_0$, the input attack signal leads to subsynchronous oscillation of the generator.
Such oscillation is easily neglected in the power system, for the generator terminal measurements stay nearly unchanged. However, the subsynchronous oscillation will cause mechanical fatigue and fracture in the long term.
In addition to the theoretical analysis in the frequency domain, we also testify the vulnerability leveraging SSR in the time domain, shown with the numerical studies in the following section.

\section{A Two-area System Example}
\label{S4}
\begin{figure}[h]
\centering
\includegraphics[width=0.9\linewidth]{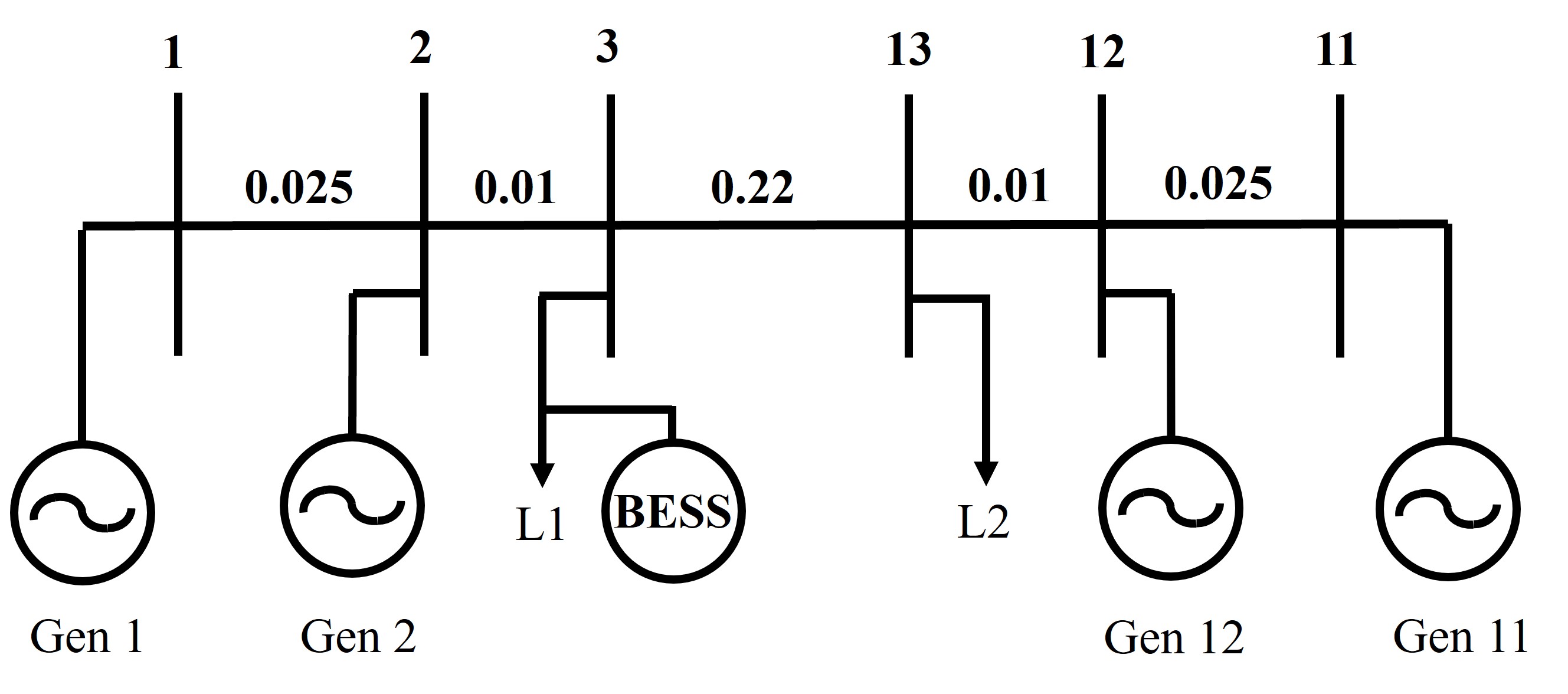} 
\caption{Two-area System}
\label{fig_sys}
\end{figure}

To investigate the existence of the vulnerability presented in this paper in the power system, we employ a commonly used two-area system for numerical studies shown in Fig. \ref{fig_sys}.

In Fig. \ref{fig_sys}, the line reactances are shown in p.u. on 230 kW and 100 MW \cite{kundur2007power}.
A linear system is considered here with a single circuit tie line.
And an energy storage device is integrated to Bus 3 to inject malicious input signal with a magnitude of 1 MW, which is 1 p.u..
The power transfer from Area 1 to Area 2 is 400 MW, where L1 = 970 MW and L2 = 1770 MW.
Four identical generating units Gen 1, Gen 2, Gen 11, and G 12, are respectively loaded to 700 MW, 670 MW, 670 MW, and 700 MW, 
with a five-mass torsional system each.
The inertia constant, power fraction, and shaft stiffness of the generator torsional system are presented in Table \ref{tab.tor}.

\begin{table}[h] 
\centering 
\caption{\small  Generator Torsional System parameters} 
\vspace{-8pt}
\includegraphics[width=1\linewidth]{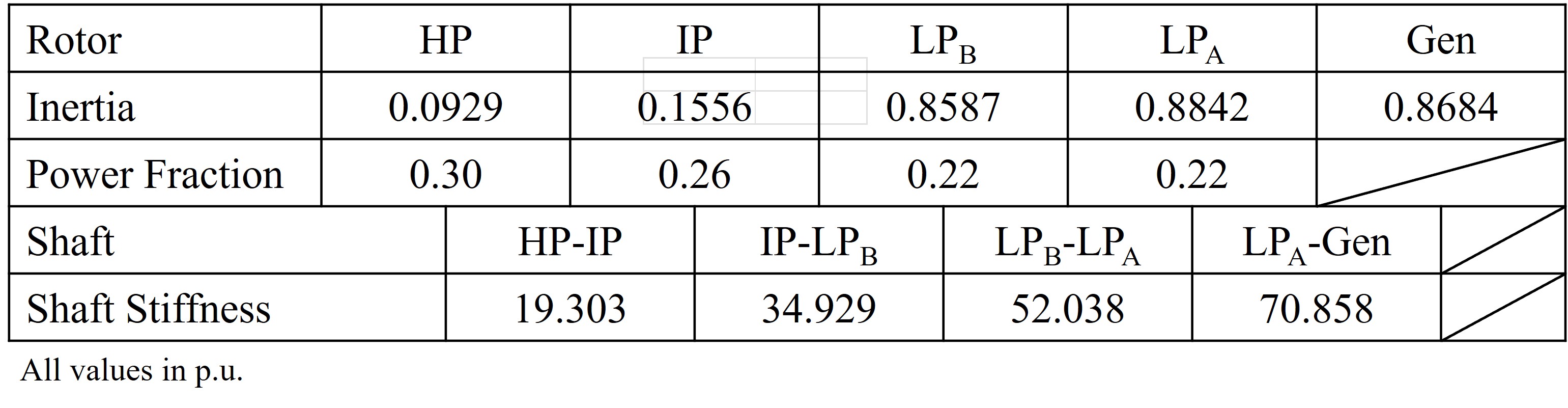}
\label{tab.tor}
\end{table}

\subsection{Frequency-Domain Analysis}
\begin{figure}[h]
\centering
\includegraphics[width=\linewidth]{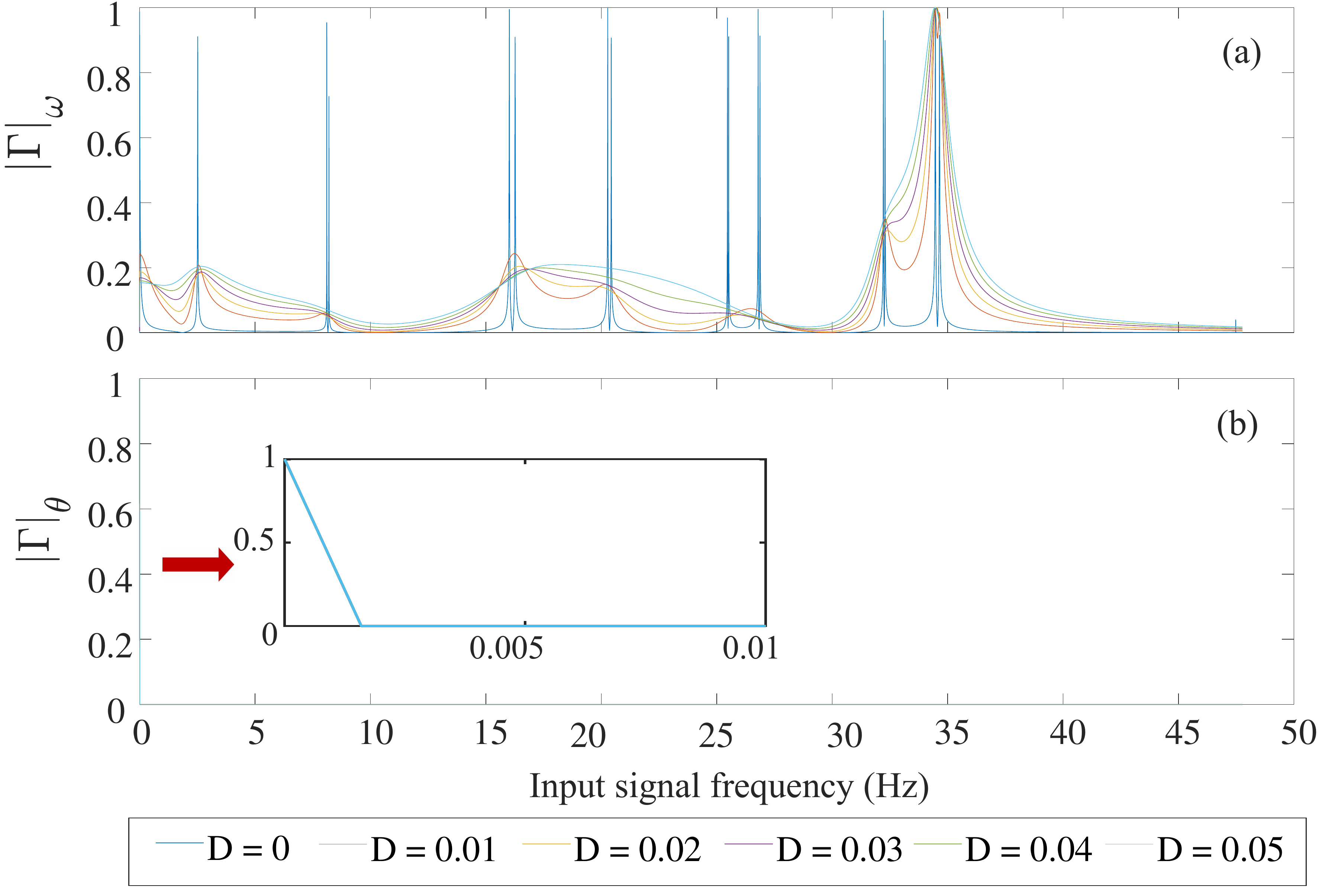} 
\caption{Transfer Function Magnitude Curves of Gen 1. }
% \todo{The bottom curve doesn't show anything?}
% \sectodo{The bottom curves have their maximums near 0, and values of the rest of the curve is insignificant comparing to the maximum. I tried to explain why the bottom figure looks like this below this figure. Or do we add a magnification to the part near 0?}
% }
\label{fig_cd}
\end{figure}

While the damping factor of the generator terminal is commonly neglected, i.e. $D_{g1}\!=\!D_{g2}\!\!=\!D_{g11}\!\!=\!D_{g12}\!=\!0$,
the damping factor of each mass in the torsional system strongly affects the transfer function magnitudes according to our tests.
The impact of the mass damping factor on the transfer function magnitude is depicted in Fig. \ref{fig_cd}.
Note that such impact on each generator is similar, we take Gen 1 as an example to avoid repeating.
In Fig. \ref{fig_cd}, the transfer function magnitude curves of rotor speed $\omega$ and angular displacement $\theta$ are respectively normalized. Thus the y-axis ranges from 0 to 1.
The $|\Gamma|_\theta$ curves in Fig. \ref{fig_cd} see their maximums with the input signal frequencies close to 0Hz.
The maximums of the $|\Gamma|_\theta$ curves are significantly higher than the other $|\Gamma|_\theta$ values, which indicates the most notable oscillation occurs when the input signals approximate DC signals.
With the increase of mass damping factor, the $|\Gamma|_\omega$ curve becomes smoother and moves upward.
A more significant oscillation is therefore expected due to the higher values of the magnitude $|\Gamma|_\omega$. 
Hence it is reasonable for us to employ a low damping factor for the torsional mass in further analysis.
A low mass damping factor is also consistent with the commonly-used steel masses in real power systems.

The values of the ratio $R_M(\omega_a)$ proposed in \eqref{rdef} are shown in Fig. \ref{fig_rm} for Gen 1:
\begin{figure}[h]
\centering
\includegraphics[width=\linewidth]{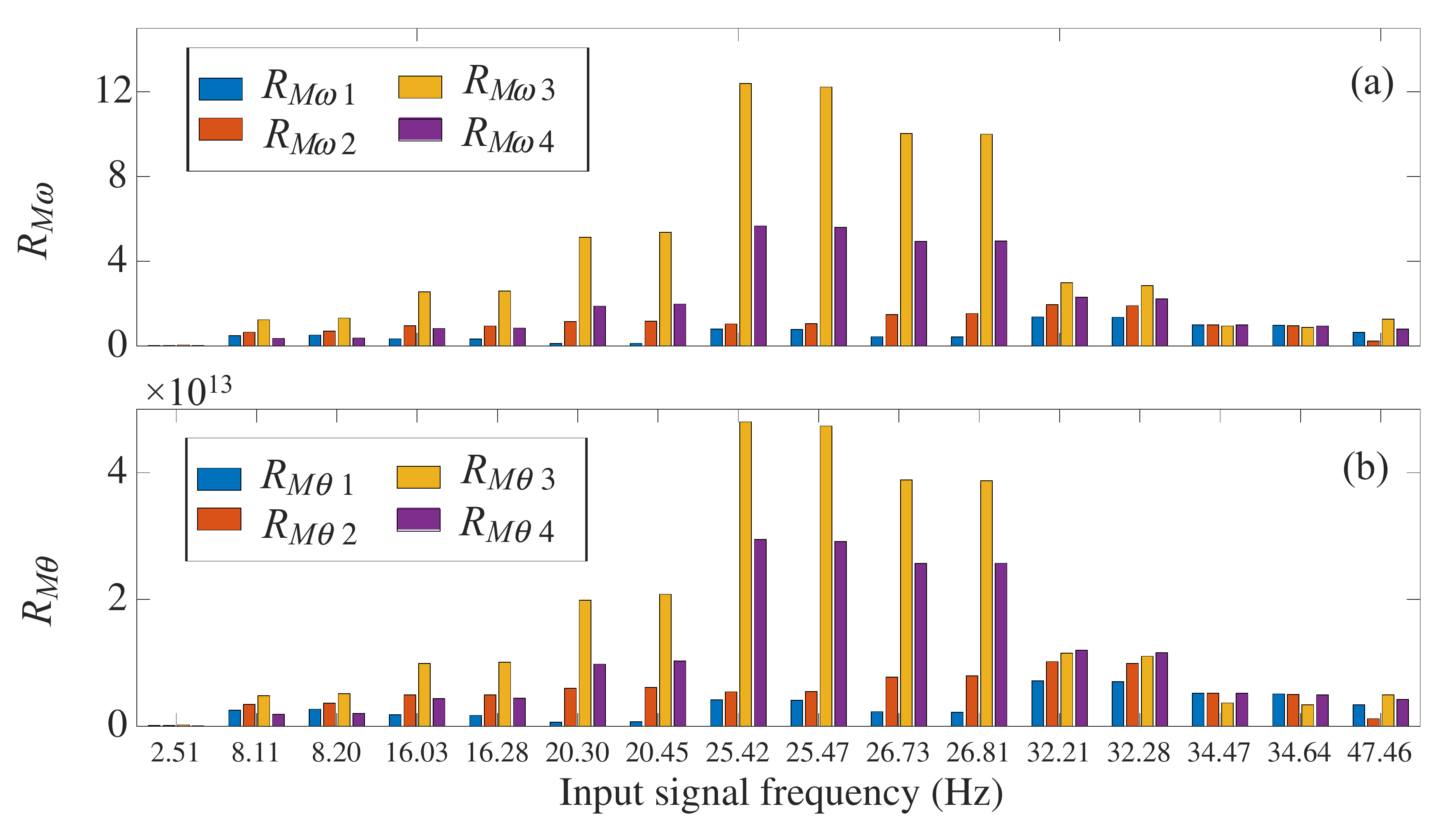} 
\caption{$R_M$ Values of Gen 1}
\label{fig_rm}
\end{figure} 
% In Fig. \ref{fig_rm}, the $R_M$ values with index $i$ are acquired though 
% $R_{M\omega i}(\omega_a)=\frac{|\Gamma_2(\omega_a)|_{\Delta \omega_{si}}}{|\Gamma_1(\omega_a)|_{\omega_{g1}}}$ for rotor speed, and
% $R_{M\theta i}(\omega_a)=\frac{|\Gamma_2(\omega_a)|_{\Delta \theta_{si}}}{|\Gamma_1(\omega_a)|_{\theta_{g1}}}$ for rotor angle.
% The x-axis indexes of Fig. \ref{fig_rm} correspond to the frequencies where local maximum values of $|\Gamma_1|_{g1}$ and $|\Gamma_2|_{g1}$.
where Fig. \ref{fig_rm} (a) presents the magnitude ratio between the torsional speed difference and the terminal rotor speed,
and Fig. \ref{fig_rm} (b) illustrates the magnitude ratio between the torsional angle difference and the terminal rotor angel.
The higher $R_M(\omega_a)$ value is, the more severe the oscillation of the torsional system will be compared to the generator terminal.
Due to the low magnitudes of their transfer functions, the generator terminal variables stay closely to their steady-state values.
The internal oscillation of the torsional system will thus be difficult to recognize.
The system is also more vulnerable when exposed to the malicious input at frequency $(\omega_a)$. 

\begin{figure}[h]
\centering
\includegraphics[width=\linewidth]{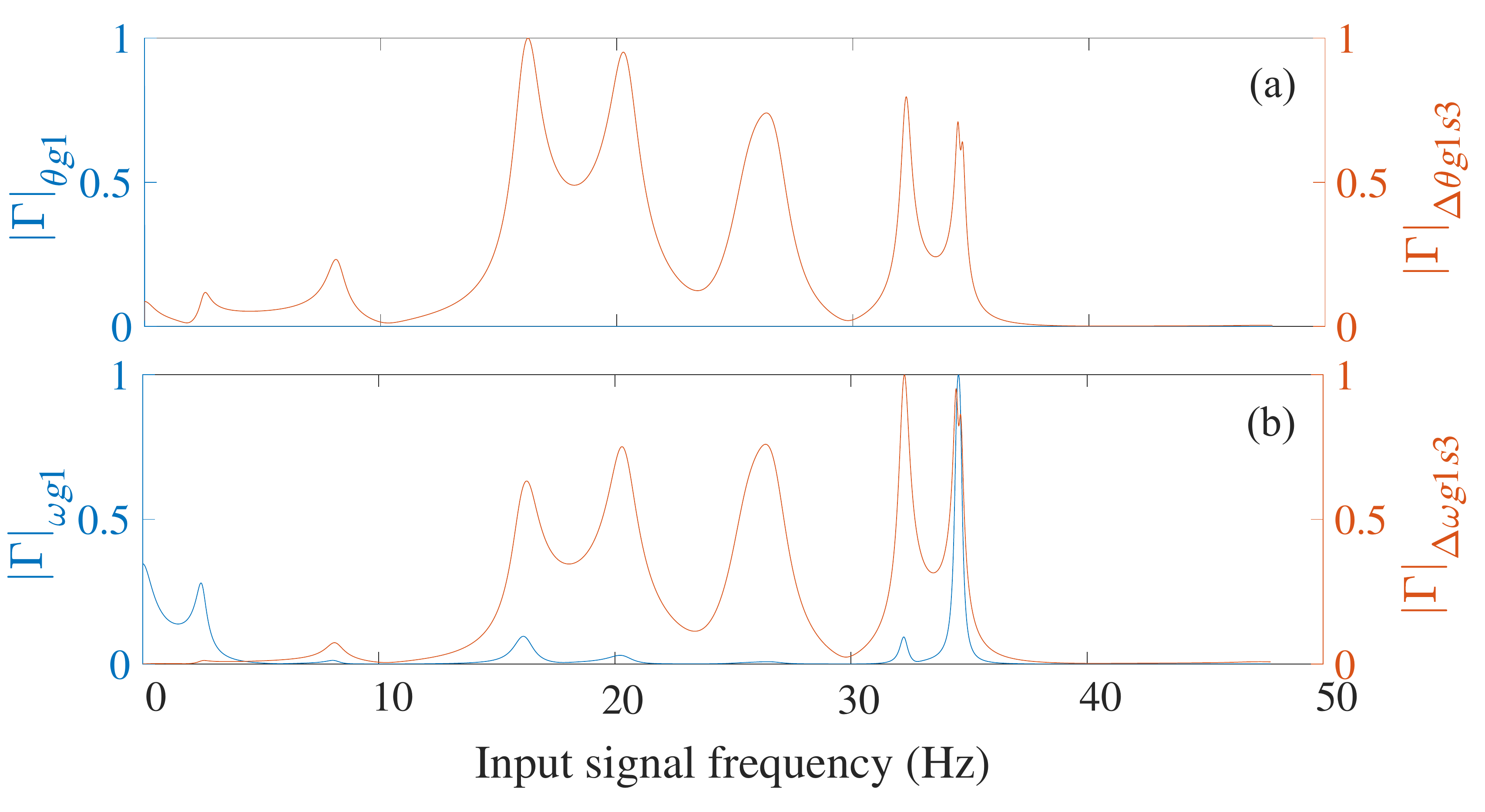} 
\caption{Comparison of $|\Gamma|_{\omega}$ ,  $|\Gamma|_{\theta}$ and $|\Gamma|_{\Delta\omega}$ ,  $|\Gamma|_{\Delta\theta}$ of Gen 1}
\label{fig_com}
\end{figure}

In Fig. \ref{fig_com}, the blue curves are the terminal speed and angle of Gen 1.
The orange curves represent the speed and angel differences between the third and fourth mass, both connecting to shaft 34 of Gen 1.

As shown in Fig. \ref{fig_com}(a), the $|\Gamma|_{\theta g 1}$ values are close to 0 when the input signal frequency is above 0Hz, which explains the reason why $R_{M\theta}$ values in Fig. \ref{fig_rm}(b) are significantly high.
Therefore the the local maximums of $R_{M\theta}$ shown in \ref{fig_rm}(b) coincide with the local maximums of $|\Gamma|_{\Delta \theta}$ shown in \ref{fig_com}(a).
In Fig. \ref{fig_com}(b), when the input signal frequency is close to 34.47Hz, both curves see a local maximum value.
This local maximum is also the global maximum of $|\Gamma|_{\theta g 1}$, meaning the oscillation at this frequency is usually damped by power system stabilizers, which are usually designed according to the speed and angle of the generator terminals.
However, the oscillations corresponding to other local maximums often remain neglected from protective measures due to the lack of attention to the speed and angle of the torsional masses.
The torsional system is therefore vulnerable to attack signals at these frequencies.
Potential attacks can thus attack the system with injection of oscillatory signals at such frequencies with an energy storage device either remotely hacked or physically controlled by them.
Considering the results shown in Fig. \ref{fig_cd}, the example system is at risk when input signal frequency lies within the range of 25.42-25.47Hz or 26.73-26.81Hz.

Note that the input signal frequency ranges depicted in Fig. \ref{fig_cd}-Fig. \ref{fig_com} are 0-50Hz, which is the typical frequency range of SSR.
From 50 to 60Hz, the y-axis values in in Fig. \ref{fig_cd}-Fig. \ref{fig_com} approximate 0, and are left out for concision. 
% Additionally, we only present the simulation results of Gen 1 in this subsection; however, the results of other generators follow the same fashion, thus are also neglected.

\subsection{Time-Domain Analysis}
Assuming the system is exposed to a square-wave input signal with magnitude $|\Delta L|= 1MW$ at Bus 3, 
we analyze the time-domain response of the output considering two attack frequencies: $f_{a1} = 25.42Hz$, and  $f_{a2} = 26.81Hz$.
The severity of an oscillation is measured by the amplitude of a signal's deviation from its steady-state value. 
We define $R_{\omega j}= \frac{max(|\Delta \omega_j-\Delta \omega_{j,0}|)}{max(|\omega_i-\Delta \omega_{i,0}|)}$,
where the numerator is the deviation of the speed difference between mass $j$ and mass $j+1$ from the initial steady-state difference.
The denominator is the deviation of the terminal rotor speed from its initial state.
Thus if the value of $R_{\omega j}$ is higher than 1, the oscillation inside the torsional system is more significant than that at the terminal rotor, and the significance increases with the rise of $R_{\omega j}$ value.
The same definition and characteristic are applied to $R_{\theta j}$, thus is not repeated.
Considering a 10s control horizon with a time step of $10^{-3}s$, an attack is issued at  $t=2s$, the values of $R_{\omega j}$ and $R_{\theta j}$ are presented in the following table.
\begin{table}[h] 
\centering 
\caption{\small $R_\omega$ and $R_\theta $ Values under Different Attack Frequencies} 
\vspace{-8pt}
\includegraphics[width=1\linewidth]{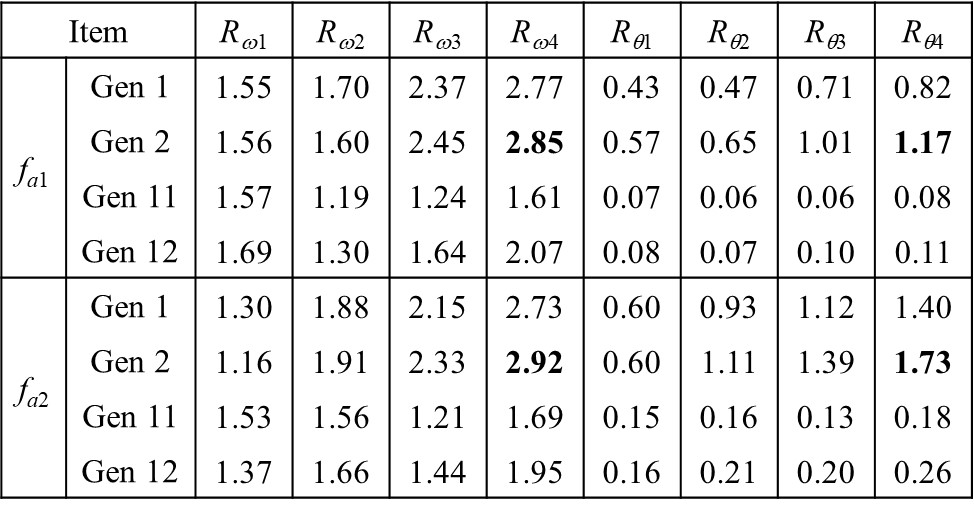}
\label{tab_tb}
\end{table}

\begin{figure}[h]
\centering
\includegraphics[width=1\linewidth]{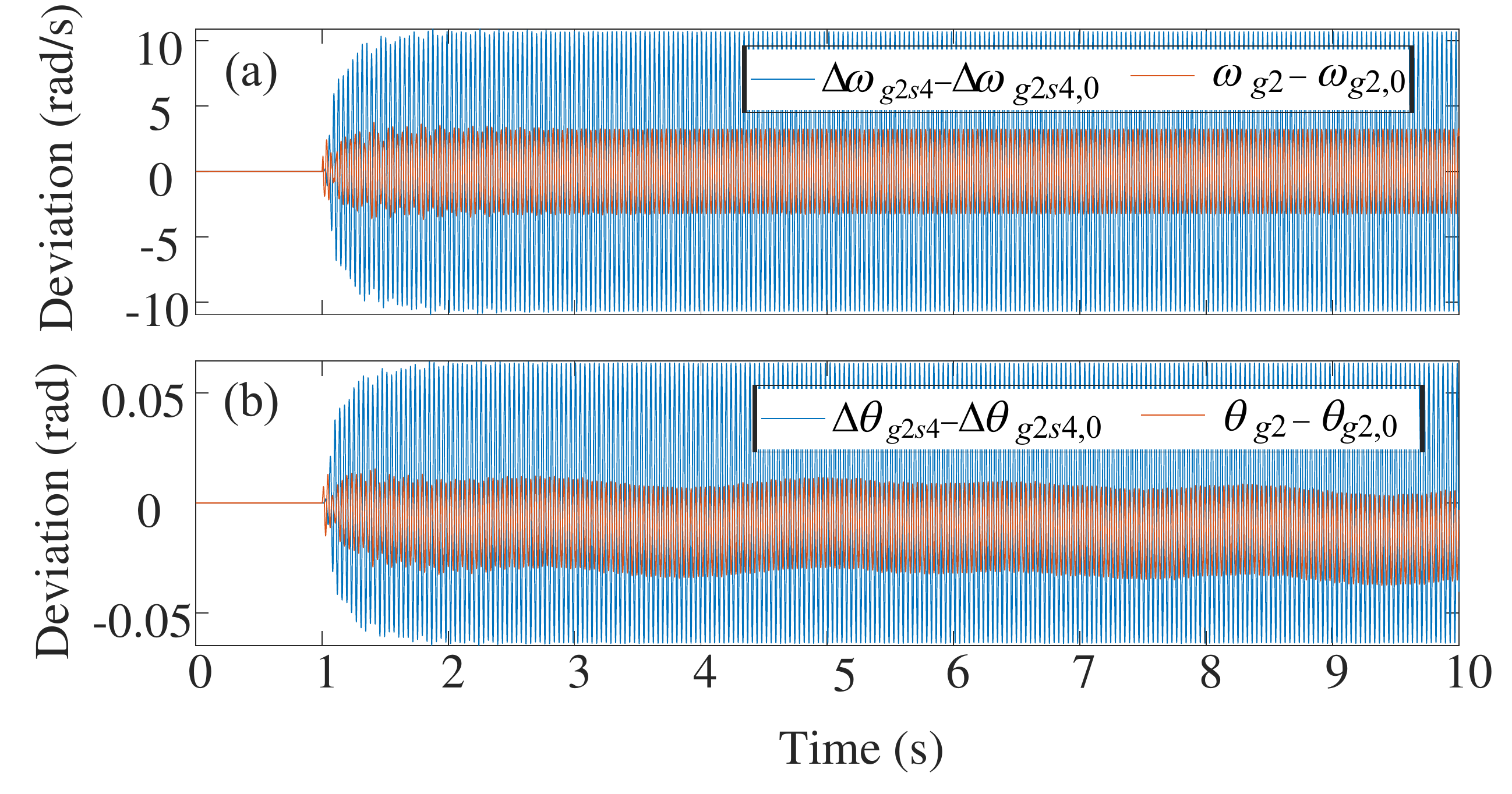} 
\caption{Time-domain Response under Attack Frequency 26.81Hz }
\label{fig_tgen}
\end{figure}

The maximums of the $R_{\omega j}$ and $R_{\theta j}$ under each attack signal frequency are highlighted in bold in Table \ref{tab_tb}.
The time-domain responses of unit Gen 2 is further illustrated in Fig. \ref{fig_tgen},
where Fig. \ref{fig_tgen}(a) depicts output $\Delta \omega_{g2s4}$ and $\omega_{g2}$,
and Fig. \ref{fig_tgen}(b) presents output $\Delta \theta_{g2s4}$ and $\theta_{g2}$.
The $R_{\omega j}$ and $R_{\theta j}$ values in Table \ref{tab_tb} higher than 1 indicate that the torsional system of the corresponding generators oscillate notably while the terminal measurements close to the initial steady-state values.
Considering attack signal frequencies $f_{a1}$ and $f_{a2}$ are both below 60Hz, the vulnerability revealed here results in SSR of the generators.
And we prove the existence of such vulnerability that leads to SSR when the system is exposed to the proposed cyber-physical attacks. 
This paper assumes access to a complete knowledge of system structure and data, which is difficult to achieve in practice. 
To explore the system vulnerability with limited access to the system information remains our future work.

\section{Countermeasures}
\label{SecAdd}

The cyber-physical attack discussed in this paper requires only a relatively low-power BESS that can be located at some distance from the targeted generators. The resulting SSR leads to metal fatigue of the generator shafts through small oscillations and that could ultimately cause a fatal mechanical failure. Further work is needed to identify generators that might be particularly susceptible to this type of attack and to develop effective countermeasures. These countermeasures can be divided into four categories: prevention, detection, reaction and mitigation.

Prevention is the first line of defense. Using best cyber security practices, it aims to prevent attackers from taking control of power devices. Unfortunately, the attack surface is getting larger due to the increasing number of BESS and other controllable components. 

Detecting this form of cyber-physical attack and locating their source are challenging issues. Because the malicious power injections are small, they are likely to be lost in the noise in conventional SCADA measurements, especially if the network is large and complex. More sensitive measurements at a higher time resolution are likely to be needed to be able to pinpoint the source of the attack, which may be in a remote part of the network.

Once an attack has been detected and its source identified, to ensure a quick reaction, procedures must be in place to disconnect it from the system. Alternatively, the targeted generators may need to be disconnected.

If prevention, detection and reaction are deemed insufficiently effective, mitigation may be required. Such measures could be similar to those that were developed to deal with naturally occurring sub-synchronous resonance. Reference \cite{4114001} groups these countermeasures into four categories: system switching and generator tripping, generator and system modifications, relaying and detecting devices, and filtering and damping devices. While some of these measures have been shown to be effective, they may be harder to implement in an adversarial context because the attack may be aimed at any generator rather than at a particular generator identified as being susceptible to sub-synchronous resonance through careful system studies. Protecting all generators against attacks at any dangerous frequency may also be difficult or very costly

\section{Conclusion}
\label{S5}
This paper exposes a potential vulnerability of power systems to a cyber-physical attack where a malicious actor could trigger a SSR in the shaft of large generators by creating small oscillations in the active power injections of a battery energy storage system. These small mechanical oscillations between the various masses connected to the shaft would cause metal fatigue and ultimately lead to a catastrophic failure of the generator. Because this resonance takes place within the shaft, it does not create significant perturbations at the terminal of the generator and might therefore be difficult to detect. Gaining access to the control system of a relatively low power BESS would therefore give an attacker an opportunity to inflict physical damage on equipment of a considerably larger rating and importance to the reliable operation of the system.
{Our further work will explore detection, reaction and mitigation countermeasures.}

% An example of a floating table. Note that, for IEEE style tables, the
% \caption command should come BEFORE the table and, given that table
% captions serve much like titles, are usually capitalized except for words
% such as a, an, and, as, at, but, by, for, in, nor, of, on, or, the, to
% and up, which are usually not capitalized unless they are the first or
% last word of the caption. Table text will default to \footnotesize as
% the IEEE normally uses this smaller font for tables.
% The \label must come after \caption as always.
%
%\begin{table}[!t]
%% increase table row spacing, adjust to taste
%\renewcommand{\arraystretch}{1.3}
% if using array.sty, it might be a good idea to tweak the value of
% \extrarowheight as needed to properly center the text within the cells
%\caption{An Example of a Table}
%\label{table_example}
%\centering
%% Some packages, such as MDW tools, offer better commands for making tables
%% than the plain LaTeX2e tabular which is used here.
%\begin{tabular}{|c||c|}
%\hline
%One & Two\\
%\hline
%Three & Four\\
%\hline
%\end{tabular}
%\end{table}


\section*{Acknowledgment}

The work described in this paper was carried out with funding from the US National Science Foundation under its CRISP - Critical Resilient Interdependent Infrastructure Systems and Processes program, grant number 1832287.



% \bibliographystyle{IEEEtran}
% \bibliography{ref}

%\documentclass[preprint,review,12pt,authoryear]{elsarticle}
% \begin{document}
%\nocite{*}
%\bibliographystyle{IEEEtran}
%\bibliography{IREP_GA}

\begin{thebibliography} {99} 

\bibitem{8672483}J. Liu, Y. Gu, L. Zha, Y. Liu, and J. Cao, “Event-triggered $H_\infty$ load
frequency control for multiarea power systems under hybrid cyber
attacks,” \textit{IEEE Transactions on Systems, Man, and Cybernetics: Systems},
vol. 49, no. 8, pp. 1665–1678, 2019.

\bibitem{8680713}  
W. Chen, D. Ding, H. Dong, and G. Wei, “Distributed resilient filtering
for power systems subject to denial-of-service attacks,” \textit{IEEE Transactions
on Systems, Man, and Cybernetics: Systems}, vol. 49, no. 8, pp.
1688–1697, 2019.

\bibitem{wang2019review}
Q. Wang, W. Tai, Y. Tang, and M. Ni, “Review of the false data injection
attack against the cyber-physical power system,” 
\textit{IET Cyber-Physical Systems: Theory \& Applications}, 
vol. 4, no. 2, pp. 101–107, 2019.

\bibitem{8260948}
J. Zhao, L. Mili, and M.Wang, “A generalized false data injection attacks
against power system nonlinear state estimator and countermeasures,”
\textit{IEEE Transactions on Power Systems}, vol. 33, no. 5, pp. 4868–4877,
2018.

\bibitem{zheng2020security}
L. Zheng, T. Gao, and X. Zhang, “Security protection and testing system
for cyber-physical based smart power grid,” in \textit{Proceedings of PURPLE
MOUNTAIN FORUM 2019-international forum on smart grid protection
and control}. Springer, 2020, pp. 847–857.

\bibitem{mousavian2017risk}
S. Mousavian, M. Erol-Kantarci, L. Wu, and T. Ortmeyer, “A risk-based
optimization model for electric vehicle infrastructure response to cyber
attacks,” \textit{IEEE Transactions on Smart Grid}, vol. 9, no. 6, pp. 6160–6169,
2017.

\bibitem{nguyen2020electric}
T. Nguyen, S. Wang, M. Alhazmi, M. Nazemi, A. Estebsari, and
P. Dehghanian, “Electric power grid resilience to cyber adversaries: State
of the art,” \textit{IEEE Access}, vol. 8, pp. 87 592–87 608, 2020.

\bibitem{zeller2011myth}
M. Zeller, “Myth or reality—does the aurora vulnerability pose a risk
to my generator?” in \textit{2011 64th Annual Conference for Protective Relay
Engineers}. IEEE, 2011, pp. 130–136.

\bibitem{assante2010high}
G. Cauley, and M. Lauby, “High-impact low-frequency event risk to the north american bulk power system,” \textit{North American Electric Reliability
Corporation (NERC), Atlanta, GA, Tech. Rep, 2010}.

\bibitem{tejada2019review}
D. A. Tejada-Arango, A. S. Siddiqui, S. Wogrin, and E. Centeno, “A
review of energy storage system legislation in the us and the european
union,” \textit{Current Sustainable/Renewable Energy Reports}, vol. 6, no. 1,
pp. 22–28, 2019.

\bibitem{glenn2016cyber}
 C. Glenn, D. Sterbentz, and A. Wright, “Cyber threat and vulnerability
analysis of the us electric sector,” Idaho National Lab.(INL), Idaho Falls,
ID (United States), Tech. Rep., 2016.

\bibitem{ieee1981proposed}
IEEE SSR Working Group, “Proposed terms and definitions for subsynchronous
resonance,” in \textit{IEEE Symposium on Countermeasures for
Subsynchronous Resonance, IEEE Pub, 81TH0086-9-PWR}, 1981, pp.
p92–97.

\bibitem{anderson1999subsynchronous}
P. M. Anderson, B. L. Agrawal, and J. E. Van Ness, \textit{Subsynchronous
resonance in power systems}. John Wiley \& Sons, 1999, vol. 9.

\bibitem{kundur2007power}
P. Kundur, “Power system stability,” \textit{Power system stability and control},vol. 10, 2007.

\bibitem{4114001}
IEEE Subsynchronous Resonance Working Group, “Countermeasures
to subsynchronous resonance problems,” \textit{IEEE Transactions on Power
Apparatus and Systems}, vol. PAS-99, no. 5, pp. 1810–1818, 1980.

\end{thebibliography}
% \end{document}

\end{document}